# Study of cluster structures of excited states for the light nuclei He, Li and Be in three-body photodisintegration processes. Application for the excited states of the $^6$He* nucleus

N. Demekhina, H. Hakobyan, J.Manukyan, A. Sirunyan and H. Vartapetian

A.I.Alikhanyan National Science Laboratory, Yerevan, Armenia

**Abstract**

The method used to investigate the cluster structures of excited states for the light nuclei He, Li and Be in three-body photodisintegration processes is described. As an application we present the experimental program on the Yerevan synchrotron for the photodisintegration of $^7$Li target, the production and decay of the $^6$He isotope and the study of the two-cluster (tritons) structures of excited states. In particular, the properties of the detectors used in the setup, as well as the precision and simulation of the measured values are presented.

**1. Introduction**

Many experimental and theoretical works have been performed to study the light nuclei structure. The structure of excited states for the light nuclei is a subject of increasing interest and is widely discussed in the modern theoretical analyses [1] that correspond to the existence and manifestation of cluster structures of excited states inside these nuclei [2, 3].

**2. Method for the study of cluster structures of nuclei's excited states**

The presented method in this work (see also [4]) concerns the study for photodisintegration process of nuclei target A= $^6$Li, $^7$Li and $^9$Be with bremsstrahlung and linear polarized beam of the Yerevan Synchrotron in the energy range $E_\gamma$ = 50 - 250 MeV (for polarized beam with energy $E_\gamma \leq 100$ MeV, the degree of polarization $P_\gamma \geq 50\%$) [5].

In this study we consider the reaction ($\gamma$ +A) with three-body final states $\gamma$+A $\to$ 1+2+3, where the three products (1, 2, 3) of the reaction are the nucleons (p, n) and also the stable particles d, t, $^3$He, $^4$He($\alpha$) (in this article we consider not only $^6$Li, $^7$Li target [4], but also A = $^9$Be and the possible emitted particle $^4$He).
In case when the three targets are $^6$Li, $^7$Li and $^9$Be and the particles 1, 2 are p, d, t, $^3$He, $\alpha$, the particle 3 is (p, n) [4].

In these conditions we observe 7 reactions of photodisintegration:

$\gamma + {}^6$Li $\to$ t + d + p
$\gamma + {}^6$Li $\to {}^3$He + d + n
$\gamma + {}^6$Li $\to \alpha$ + p + n
$\gamma + {}^7$Li $\to$ t + t + p
$\gamma + {}^7$Li $\to {}^3$He+ t + n
$\gamma + {}^7$Li $\to \alpha$ + d + n
$\gamma + {}^9$Be $\to \alpha$ + $\alpha$ + n

There are two types of channels for disintegration of the excited states of isotope-targets:
- A statistical channel γ +A→1+2+3 which represents a physical background channel of the reaction (γ +A) and which can be calculated as three-body phase-space.
- The channels of production and decay of the excited isotope states in the given reaction γ +A→ 1+2+3 and precisely in our case in the form of three two-body processes: γ +A→ (12)+N, γ +A→ (1N)+2 and γ +A→ (2N)+1 with 3 resonances' excited states: B*=(12)*, (1N)*, (2N)*. They are decaying in the final three-particle states.

For the study of resonances as (12)* (in the process γ +A→ (12)+N) with the two products [(1N)*+2] and [(2N)*+1] also present, it is necessary to determine the contribution of these two resonances (1N)*, (2N)* in the studied process.

In case of the excited states isotope $^6$He (t+t) (reaction γ + $^7$Li → t + t + p), and on the question of the contribution of resonance states $^4$He (t+p) in this reaction see below.

For the study of reactions γ +A→ 1+2+3, the production angles and the kinetic energy of the particles (1, 2) are measured in coincidence in two detectors of the setup.

These measurements are able to determine the energy ($E_\gamma$) of incident photon on the target, effective mass ($M_{12}$) of excited states of two clusters (1, 2), the parameters of the unregistered particle (nucleon N) and the Dalitz diagram of the studied reaction[4].

For the known (or to be determined by our method also [4], in the future) excited states of seven isotopes $^5$He, $^6$He, $^5$Li, $^6$Li, $^7$Li, $^8$Be and $^9$Be it is possible to determine the 12 cluster structures of excited isotope states (see below the case of isotope $^6$He*(t+t)) by the method presented in this work (also using polarized photon beam [5]):

1- $^5$He*(t+d)
2- $^6$He*(t+t) and $^6$He*($^4$H*+d) by another methods[3]
3- $^5$Li*($^3$He+d) and ($^4$He+p)
4- $^6$Li*($^3$He+t), ($^4$He+d), ($^5$He*+p) and ($^5$Li*+n)
5- $^7$Li*($^6$He*+p) and ($^6$Li*+n)
6- $^8$Be (α+α)
7- $^9$Be ($^8$Be*+n)

### 3. Application for the two-cluster(t+t) excited states of $^6$He* nucleus

The results of experimental studies of two-cluster (t+t) excited states of $^6$He* nucleus obtained so far by two various methods [6],[7] differ, concerning the number of excited states of $^6$He*, as well as the values of the energies and widths.

For reaction $^7$Li+ $^6$Li→ $^7$Be+ ($^6$He*(t+t)) the results are:

E=18,0±0.5MeV, Γ=7.7±1.1MeV [6(1)]
E=18,0±1.0MeV, Γ=9.5±1.0MeV [6(2)]
For reaction $\pi^- + {}^9Be \rightarrow {}^6He^*(t+t)+t$ [7] the results are:
$E_a$=15.8±0.6MeV, Γ=1.0±0.6MeV ( a )
$E_b$=20.9±0.3MeV, Γ=3.2±0.5MeV ( b )
$E_c$=31.1±1.0MeV, Γ=6.9±2.3MeV ( c )

At first, by the presented method we propose the study of two-cluster structure of excited states of $^6$He in the process of $^7$Li nuclei photodisintegration using bremsstrahlung and linearly polarized photon beams [5] of the Yerevan synchrotron in the energy range 50-250MeV.

It is expected that in the considered reaction $\gamma + {}^7Li$ with the excited states $^6$He*(t + t) +p the spectrum of excited states of $^6$He will be close to the one measured in [7], where the final excited states in the reaction $\pi^- + {}^9Be$ are $^6$He*(t+t)+t , whereas in the reaction $^7Li+{}^6Li$, the final excited state is $^7Be+ {}^6He^*(t+t)$ [6(1,2)].

The performance of our experimental setup allows to achieve the necessary precision, close to the one of experiment [7] for measurements of $^6$He*(t+t) excited states in the energy range of 15-30MeV .

For this purpose we consider the process of photodisintegration of $^7$Li target with bremsstrahlung photons in the three-particle final state reaction $\gamma + {}^7Li \rightarrow ({}^6He^*+p) \rightarrow t + t + p$ and the detection of the coincidence of the two produced triton particles.

## 3.1. Layout of the experimental setup

The scheme of the photodisintegration experiment is shown in Fig. 1. The beam of bremsstrahlung photons, generated by electrons on the synchrotron`s internal target, is collimated (K) and cleaned (ML) from the charged particles. The γ-beam intensity is measured by Wilson- type quantametre.

The particular configuration presented below is adapted to the average photon energy of 75 MeV. Experimental setup registers two tritons in coincidence and it consists of two telescopes of silicon detectors, allowing to identify the particles' type, measure their kinetic energy and production angles, as well as completely restore the kinematics of three-body reaction.

Each telescope consists of two thin (dE/dx), 2 mm strip size Si sensors and single thick (E) Si detector. The first detector has horizontal strips and measures the horizontal coordinates and the second – vertical coordinates, that allows to define polar and azimuthal angles of emitted tritons with good accuracy. The choice of thicknesses for lithium target (200 μm) and Si detectors (50,150 and 1000 μm respectively) allows to measure the tritons' kinetic energy in the range of $E_t$ = 4.5 -18 MeV. Geometrically the telescopes are located at a distance of 20 cm from a target, covering a solid angle app.0.1 sr each.

## 3.2. Monte Carlo simulation of photodisintegration experiment $\gamma + {}^7Li \rightarrow t + t + p$

To determine the kinematical distributions, to elaborate the requirements to the experimental setup for the measurements' accuracy of the photon energy and effective masses of the two tritons, as well as to maximize the yields of investigated process, the Monte Carlo calculations were carried out for the average energy $E_\gamma$ = 75 MeV of photon beam.

The phase-space simulation programs [8] completed with the user code [5] allowed to describe the details of setup's geometry, including the shape of photon spectrum, the size of the beam in the target, multiple scattering and energy loss of particles inside the target and detectors. The programs were used to calculate kinematic distributions in three-body $\gamma + {}^7Li \rightarrow t + t + p$ and quasi-two-body $\gamma + {}^7Li \rightarrow {}^6He^* + p$ processes with subsequent decay of the excited ${}^6He^*$ state (${}^6He^* \rightarrow t + t$) at a given photon energy. In calculations, the experimental data on the energies and widths of three levels of excited ${}^6He$ [7] were used.

As an illustration the Fig.2 shows the angular distributions of two tritons in coincidence for energy level (b) of the excited ${}^6He^*$ at the photon energy $E_\gamma$ = 75 MeV. As it is seen the optimal polar-angle setting for two detectors is identical ($\theta$ = 75$^0$), while the azimuthal angular difference ($\varphi_1$-$\varphi_2$=100$^0$) reflects the non-collinearity of two tritons' tracks in three-body kinematics. The Monte-Carlo simulation and subsequent reconstruction of kinematics allow to define an experimental resolution for photon energy (see Fig. 3) confirming the choice of silicon detector granularity (2 mm) and tritons' kinetic energy resolution ($\sigma/E_t$ =1%), as well as allow to define the photon energy with accuracy of $\sigma_\gamma/E_\gamma$ = 0.8 % within an aperture of the telescopes' solid angles.

The basic requirement of the planned experiment is to provide a necessary accuracy of determination at various levels of excited ${}^6He^*$ nuclei and corresponding cross-sections. For this purpose the analysis of three-body process data using the effective masses spectrum of two tritons is most straightforward. The full Monte-Carlo simulation and subsequent reconstruction of kinematics allow to define an experimental resolution of the effective masses $M_{12}$.

From the calculated effective mass distributions of two tritons one can get data of excited ${}^6He^*$ isotope energy, that is defined as $E_x = M_{12} - m_{He}$, where $m_{He}$ is the mass of ${}^6He$. Obtained results for detector's energy resolution in case of level (b) [7], with confirmed silicon sensor granularity and kinetic energy resolution is shown in Fig.4. As is seen from the figure, the accuracy of energy determination of excited state ${}^6He^*$ at $E_\gamma$ = 75 MeV doesn't exceed $\sigma/E_x$ =0.5 %.

The Dalitz diagram also allows to select the analysed phase-space with the visible or expected resonant structure inside the two-triton system (${}^6He^* \rightarrow t+t$), as well as for the proton and triton (${}^4He^* \rightarrow t+p$), that corresponds to different channels of ${}^7Li$ photodisintegration with identical final states. The data of excited nuclei ${}^4He^*$ levels are taken from compilation [9].

As an illustration of this analysis, Fig.5 shows the Dalitz plots for two tritons registered in coincidence, where $T_1$ and $T_2$ correspond to the kinetic energy in the CM system. As it is seen from the figure, the levels of ${}^6He^*$ excitation are well-separated from each other, as well as from the events of ${}^4He^*$ excitation.

It should be noted that for experimental study of ${}^7Li$ photodisintergration process with three-body final states we will continue the Monte Carlo calculations.

**4. Conclusion**

Within the presented proposal for investigation of three-body photodisintegration processes of the light nuclei on an example of $^7$Li, we plan to study with bremsstrahlung beam of the Yerevan synchrotron the cluster structures of excited states of seven isotopes $^5$He, $^6$He, $^5$Li, $^6$Li, $^7$Li, $^8$Be and $^9$Be in the energy range $E_\gamma$ = 50 - 250 MeV, including polarized photon beam.

It is shown, that the presented method to study the cluster structures of isotope $^6$He* at average energy $E_\gamma$ = 75 MeV photon beam on the basis of registration decayed tritons in coincidence by the two telescopes of silicon detectors, allows separate energy levels of excited isotope $^6$He* with an accuracy of 0.5 % and determination of photon energy with precision of $\sigma_\gamma / E_\gamma$ = 0.8 % .

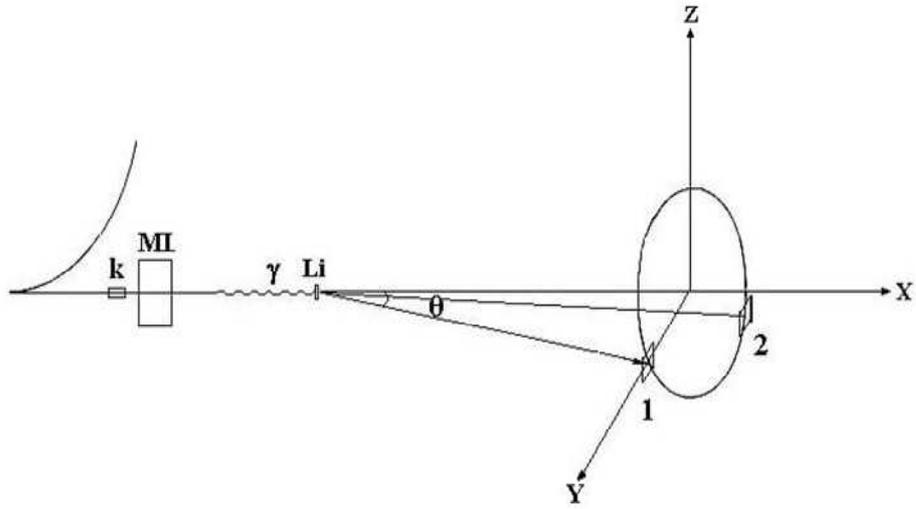

Fig.1

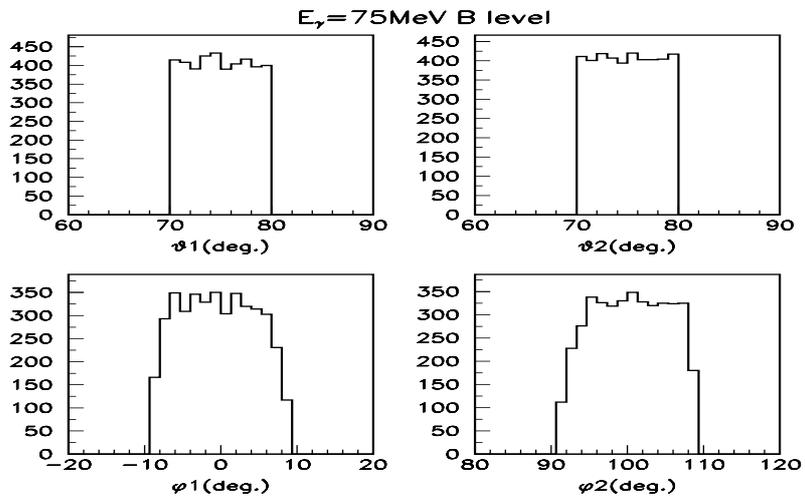

Fig.2

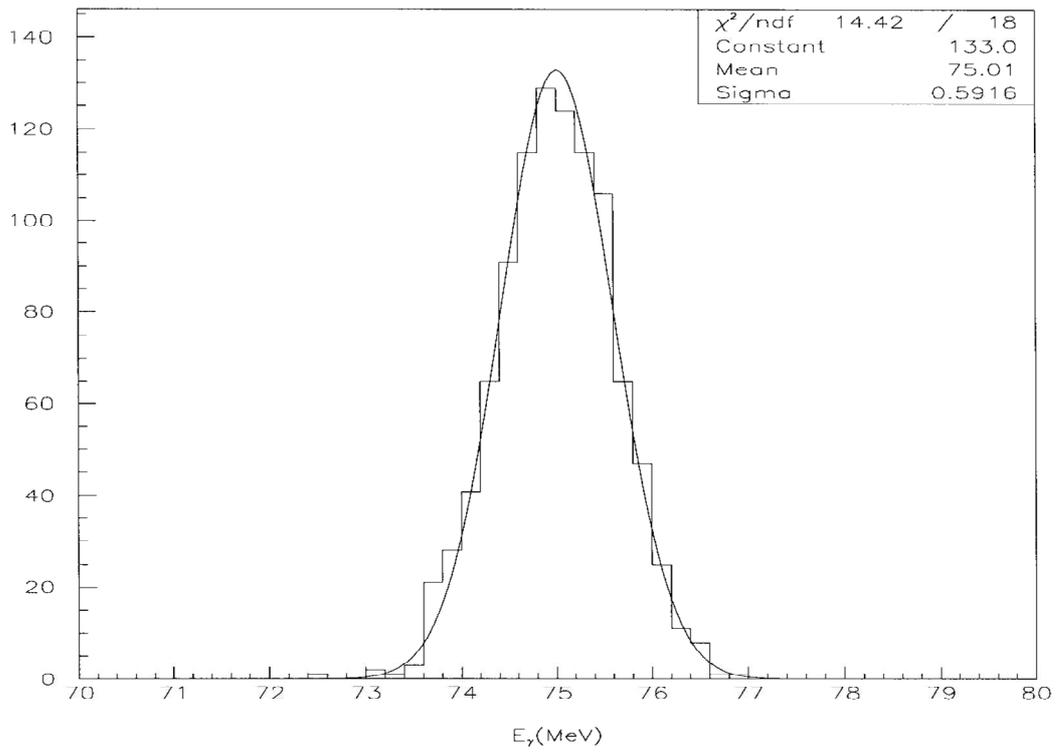

Fig.3

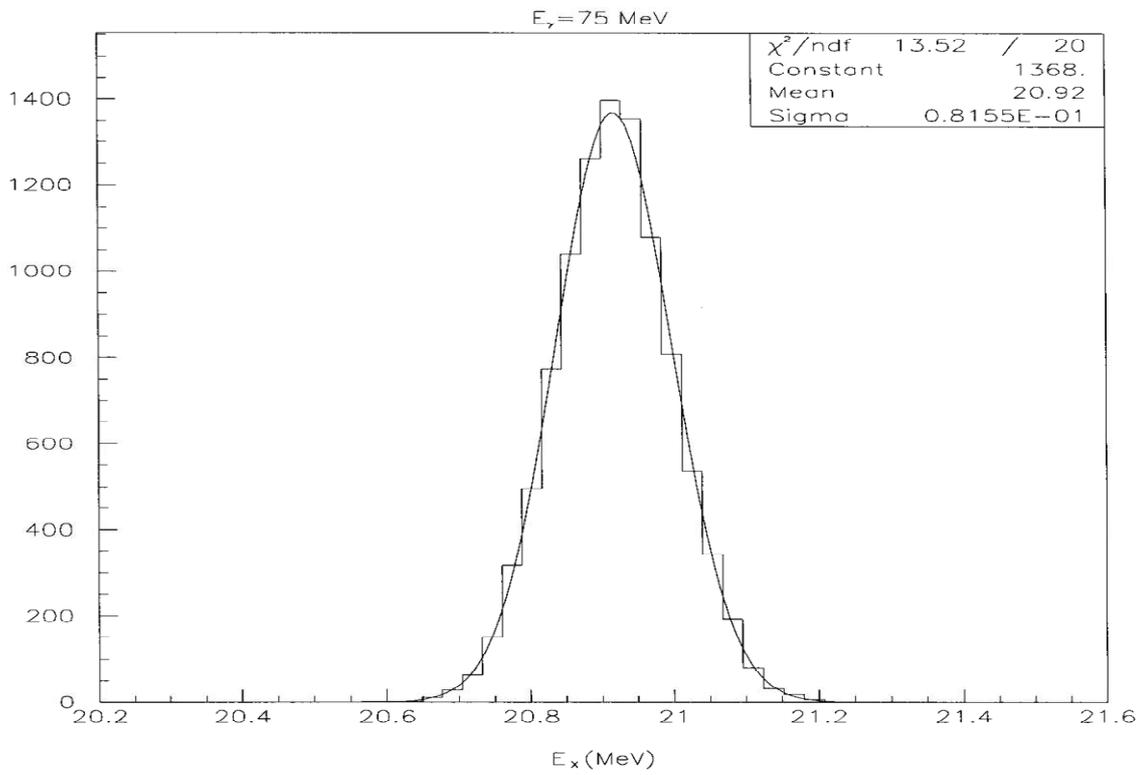

Fig.4

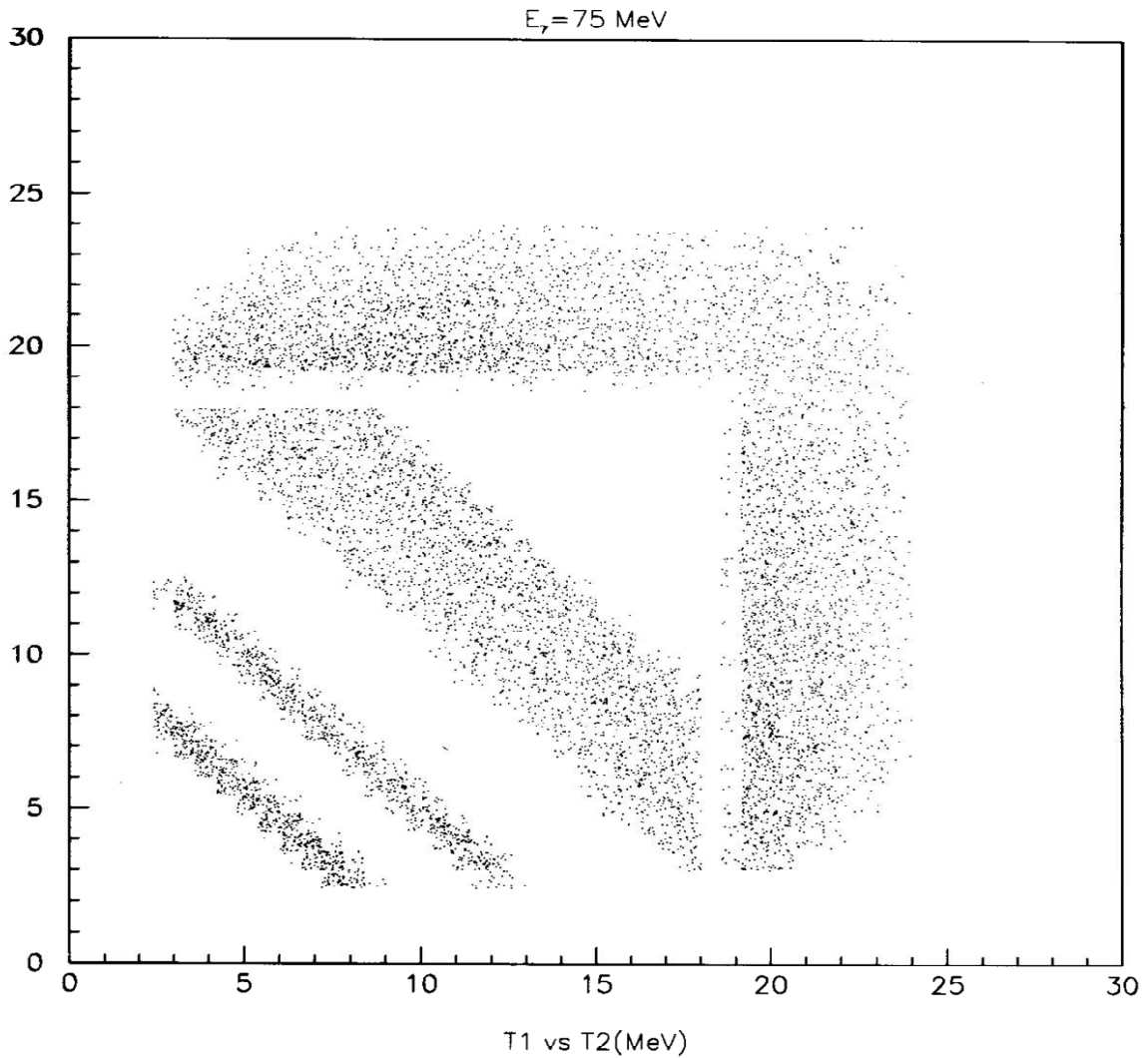

Fig.5

**Figure captions**

Fig.1. Scheme of the experimental setup.

Fig.2. Polar and azimuthal angular distributions of two tritons from excited isotope $^6$He* detected in telescopes at the photon energy $E_\gamma = 75$ MeV.

Fig.3. Photon energy distribution of two tritons detected in coincidence confirming silicon detector granularity and its energy resolution.

Fig.4. Experimental resolution for the He* excitation energy for two tritons.

Fig.5. The Dalitz plots for two tritons registered in coincidence for two different channels of $^7$Li photodisintegration ($^6$He*+ p) and ($^4$He*+ t), which corresponds to three levels (a,b,c [7]) and level $E_x$= 28.3 ± 9.8 MeV [9], respectively .